\documentclass[12pt,a4paper]{article}

\usepackage{amsmath}
\usepackage{amssymb}
\usepackage{amsfonts}
\usepackage{dsfont} % for unit operator  (indicator function) \mathds{1}
\usepackage[usenames,dvipsnames]{color}

\usepackage{graphicx}
\newcommand{\be}{\begin{equation}}
\newcommand{\ee}{\end{equation}}
\newcommand{\bel}[1]{\begin{equation}\label{#1}}
\newcommand{\bea}{\begin{eqnarray}}
\newcommand{\eea}{\end{eqnarray}}
\newcommand{\balign}{\begin{align}}
\newcommand{\ealign}{\end{align}}
\newcommand{\ba}{\begin{array}}
\newcommand{\ea}{\end{array}}
\newcommand{\bfig}{\begin{figure}}
\newcommand{\efig}{\end{figure}}

\newcommand{\eref}[1]{(\ref{#1})}

\newcommand{\Fref}[1]{Figure~\ref{#1}}

\allowdisplaybreaks

\newcommand{\bra}[1]{\mbox{$\langle \, {#1}\, |$}}
\newcommand{\ket}[1]{\mbox{$| \, {#1}\, \rangle$}}
\newcommand{\exval}[1]{\mbox{$\langle \, {#1}\, \rangle$}}
\newcommand{\inprod}[2]{\mbox{$\langle \, {#1} \, | \, {#2} \, \rangle$}}
\newcommand{\iz}{\, |\,}
\newcommand{\Prob}[1]{\mbox{${\rm Prob}\left[ \, {#1}\, \right]$}}

\newcommand{\ot}{\otimes}

\newcommand{\rmd}{\mathrm{d}}
\newcommand{\rme}{\mathrm{e}}

\newcommand{\C}{{\mathbb C}}

\newcommand{\Z}{{\mathbb Z}}

\newcommand{\T}{{\mathbb T}}

\newtheorem{theo}{Theorem}[section]
\newtheorem{lmm}[theo]{Lemma}
\newtheorem{df}[theo]{Definition}

\newtheorem{cor}[theo]{Corollary}

\def\qed{\hfill$\Box$\par\medskip\par\relax}

\begin{document}

\title{Microscopic structure of shocks and antishocks in the ASEP conditioned on low current}
\author{V.~Belitsky$^{1}$
\and G.M.~Sch\"utz$^{2,3}$ 
}

%\begin{document}

\maketitle

{\footnotesize
\noindent $^{~1}$Instituto de Matem\'atica e Est\'atistica,
Universidade de S\~ao Paulo, Rua do Mat\~ao, 1010, CEP 05508-090,
S\~ao Paulo - SP, Brazil
\\
\noindent Email: belitsky@ime.usp.br
%\noindent url: \texttt{http://www.proba.jussieu.fr/$\sim$comets}

\smallskip
\noindent $^{~2}$Institute of Complex Systems,
Forschungszentrum J\"ulich, 52425 J\"ulich, Germany
\\
\noindent Email: g.schuetz@fz-juelich.de

\smallskip
\noindent $^{~3}$Interdisziplin\"ares Zentrum f\"ur komplexe Systeme, Universit\"at
Bonn, Br\"uhler Str. 7, 53119 Bonn, Germany\\
\noindent URL: http://www.izks.uni-bonn.de
}

\begin{abstract}
We study the time evolution of the ASEP on a one-dimensional torus with $L$ sites, conditioned 
on an atypically low current up to a finite time $t$. For a certain one-parameter family of initial measures with 
a shock we prove that the shock position performs a biased random walk on the torus and that the measure
seen from the shock position remains invariant. We compute explicitly the transition rates of the random walk. 
For the large scale behaviour this result suggests that there is an atypically low current such that the 
optimal density profile that realizes this current is a hyperbolic tangent with a travelling shock discontinuity.
For an atypically low local current across a single bond of the torus we prove that
a product measure with a shock at an arbitrary position and an antishock at the conditioned bond
remains a convex combination of such measures at all times which implies that the antishock 
remains microscopically stable under the locally conditioned dynamics.
We compute the coefficients of the convex combinations.
 \\[.3cm]\textbf{Keywords:} Asymmetric symmetric exclusion process; large deviations; shocks; antishocks.
\\[.3cm]\textbf{AMS 2010 subject classifications:}
82C22, Secondary: 60F10, 60K35.

\end{abstract}

\newpage

\section{Introduction}
\label{s_intro}

The Asymmetric Simple Exclusion Process (ASEP) on $\Z$ \cite{Ligg99,Schu01}
describes the Markovian time evolution of identical particles on $\Z$ according to 
the following rules:
{\it (1)}\ Each particle performs a continuous time simple random walk on
with hopping rate $c_r$ to the right  and $c_\ell$ to the left; 
{\it (2)}\ The hopping attempt is rejected when the
site where the particles tries to move is occupied.
Without loss of generality $c_r>c_\ell$ is assumed. An important
family of invariant measures for this process are the Bernoulli product
measures parametrized by the particle density $\rho$. The stationary current
$j^\ast$ is then $j^\ast = (c_r-c_\ell) \rho(1-\rho)$. For $c_r=c_\ell$ (symmetric 
simple exclusion process) the dynamics is reversible and one has $j^\ast=0$. 
The process may also be defined on a finite integer lattice with various types of 
boundary conditions which may lead to non-vanishing stationary currents  
even in the symmetric case. The presence of a stationary current
indicates the lack of reversibility of the dynamics and therefore one is interested
not only with its stationary expectation $j^\ast$, but also its fluctuations.

%Its rich stationary and dynamical behaviour, its exact solvability and its many
%applications as a toy model for non-equilibrium behaviour, have made the ASEP
%the best-studied interacting particle system both in mathematical probability
%theory and in non-equilibrium statistical physics \cite{Ligg99,Schu01,Chow10}.
Starting with the seminal papers \cite{Derr02,Bert02} the large deviation theory
for the ASEP has been developed in considerable detail. 
Of particular interest in this context is the time-integrated
current $J_k(t)=J^+_k(t) - J^-_k(t)$ across a bond $(k,k+1)$, where $J^+_k(t)$ is
the number of jumps of particles from $k$ to $k+1$ up to time $t$ and analogously
$J^-_k(t)$ is the number of jumps from $k+1$ to $k$ up to time $t$, starting from some
initial distribution of the particles. In a periodic system
with $L$ sites, i.e., on the torus $\T_L:=\Z/L\Z$, a related quantity of interest is the 
time-integrated total
current $J(t) =  \sum_{k\in\T_L} J_k(t)$ which is intimately related to the rate of 
entropy production \cite{Lebo99,Harr07}. We denote its
distribution by $Z_J(t) := \Prob{J(t) = J}$ with $J\in\Z$.
One also consideres the (local) mean current $j_k(t) = J_k(t)/t$ and the
(global) mean current density $j(t) = J(t)/(Lt)$. For the
Bernoulli product measure one has $\lim_{t\to\infty} j_k(t) = \lim_{t\to\infty} j(t)= j^\ast $. 
The probability to observe for a long time interval
$t$ an untypical mean $j\neq j^\ast$ is exponentially small in $t$. This is expressed in the
large deviation property \cite{Derr98}
$Z_J(t) \propto \exp{(-f(j)Lt)}$ where $f(j)$ is the rate
function which plays a role analogous to the free energy in equilibrium statistical
mechanics. Indeed, in complete analogy to equilibrium one 
introduces a generalized fugacity $y=\rme^s$ with generalized
chemical potential $s$ and also studies the generating
function $Y_s(t) = \exval{y^{J(t)}} = \sum_J y^JZ_J(t)$. 
The cumulant function $g(s) = \lim_{t\to\infty} \ln{Y_s(t)}/(Lt)$ is the 
Legendre transform of the rate function for the mean current, $g(s) = \max_j [ j s - f(j)] $.
The intensive variable $s$ is thus conjugate to the mean current density $j$. 

Recently a focus of attention has been on the spatio-temporal structure of the process
conditioned on realizing a prolonged untypical behaviour of the current. 
A convenient approach to study this rare and extreme behaviour is to consider the process in terms of
the conjugate variable $s$. Fixing some $s\neq0$ corresponds to studying realizations of the process where
the current fluctuates around some non-typical mean. We shall refer to this approach as 
grandcanonical conditioning, as opposed to a canonical condition where the current $J(t)$ would 
be conditioned to have some fixed value $J$. For the weakly asymmetric 
exclusion process (WASEP) on $\T_L$ where $ c_r-c_\ell = \nu/L$ is small,
it turns out in the hydrodynamic limit that for
any strictly positive $s$ (i.e., for any current $j>j^\ast$) the optimal macroscopic density
profile $\rho(x,t)$ that realizes such a deviation 
is time-independent and flat \cite{Bodi05}.  Thus on macroscopic scale the conditioned density
profile is equal to the typical (unconditioned) density profile. However,
a recent {\it microscopic} approach using a form of 
Doob's $h$-transform has 
revealed \cite{Popk10,Popk11} that for large $s$ the ASEP with arbitrary strength of the 
asymmetry $ c_r-c_\ell$
exhibits interesting stationary correlations. Even more 
remarkably, the process undergoes a phase transition from the typical dynamics 
in the KPZ universality class with dynamical exponent
$z=3/2$ to ballistic dynamics with $z=1$. 

For currents {\it below} the typical value ($s<0$) the large deviation approach for the WASEP yields
a different phenomenon \cite{Bodi05}: For  currents close to the typical current (but $j<j^\ast$) 
the optimal macroscopic density profile $\rho(x,t)$ 
is time-independent and flat as is the case for any $s>0$.
However, below some critical value $s_c$ (i.e. below some critical $j_c<j^\ast$)
the flat profile becomes
unstable and a travelling wave of the form $\rho(x-vt)$ develops.
In the limit $\nu \to\infty$ (expected to correspond to finite asymmetry in the ASEP) 
the optimal profile in this regime is predicted to have the form of a step
function with two constant values $\rho_1$ and $\rho_2$.
This is a profile consisting of a shock discontinuity where the density
jumps from $\rho_1$ to $\rho_2>\rho_1$ at a position
$x_1(t)$ and an antishock  where the density
jumps from $\rho_2$ to $\rho_1$ at position
$x_2(t)$. Hence, conditioning on a sufficiently large
negative deviation of the current $j < j_c$ induces a phase separation in the WASEP into a low density
segment of length $r=x_1-x_2$ with density $\rho_1$ and a high density segment
of length $L-r=x_2-x_1$ with density $\rho_2$ (positions and distances taken modulo $L$).
A similar phase separated optimal profile, but without drift, has recently been
obtained for atypically low activities $K(t) = J^+(t) + J^-(t)$ in the SSEP
\cite{Leco12}.

In order to understand this phenomenon better we
develop in this paper a microscopic approach for the ASEP for atypically low currents, 
but without restriction to the case of weak asymmetry. 
For a periodic system, 
as discussed above, one may expect in the long-time regime 
a travelling wave. In order to get insight into the microscopic
structure of the travelling wave
we focus on short times and address 
the question how for global conditioning a certain initial distribution which already has a shock
evolves at some later {\it finite} time $t$. We also consider local conditioning of the current
on a single bond. For this case
we study the evolution for finite times $t$ of a shock/antishock pair 
with densities $\rho_{1,2}$.
Our rigorous analysis is an adaptation of the algebraic techniques developed in
\cite{Beli02} and extended recently by Imamura and Sasamoto to study
current moments of the ASEP \cite{Imam11}. This approach -- which has its probabilistic origin 
in the self-duality of the ASEP \cite{Schu87,Giar09,Boro12} --
constrains our approach to a specific family of initial shock measures which are certain
functions of the hopping asymmetry. However, going from the
weakly asymmetric case all the way to the totally asymmetric limit allows us to cover a wide range
of scenarios.

The paper is organized as follows. In Sec. 2 we describe informally the well-established, but not
so widely known tools required for studying grandcanonically conditioned dynamics.
In Sec. 3 we introduce a new family of shock measures which are relevant for the
grandcanonically  conditioned time evolution of the ASEP.
In Sec. 4 we state and prove the main result and in Sec. 5 we conclude with some further comments
 on a generalization of the present approach to more general shock measures and 
on the macroscopic large deviation results of \cite{Bodi05}.

\section{ASEP conditioned on a low current}
\label{lowcurrent}

We consider the ASEP on the one-dimensional torus $\T_L$  with $L$ sites.  For convenience 
we take $L$ even and denote sites (defined modulo $L$) by an integer in the interval $-M+1,M$
where $M:=L/2$. We shall identify jumps to the ``right'' with clockwise jumps. 
We introduce the hopping asymmetry
\bel{asymmetry}
q = \sqrt{\frac{c_r}{c_\ell}}
\ee
and parametrize the densities $\rho$ by the
fugacity
\bel{fugacity}
z := \frac{\rho}{1-\rho}.
\ee
The shock/antishock initial distributions considered below are characterized by two fugacities
$z_{1,2}$ satisfying\\

\noindent {\bf Condition S:} 
\bel{central}
\frac{z_2}{z_1}=q^2
\ee
By convention we shall assume $1<q<\infty$, i.e. we consider preferred 
hopping to the ``right'' (clockwise) and we exclude the symmetric case $q=1$.

Grandcanonically conditioned Markov processes may be studied in the spirit of Doob's $h$-transform
\cite{Doob}, as was done for the ASEP conditioned on very large currents 
in \cite{Simo09,Popk10,Popk11}, see below for the precise definition. 
To this end, we follow our earlier work \cite{Beli02} and
we employ a matrix formulation to define the process. This formulation of Markovian dynamics,
formalized in probabilistic terms for exclusion processes in \cite{Lloy96} and described 
informally in detail in \cite{Schu01}, allows
for an alternative convenient formulation particularly of conditioned interacting particle 
systems, as discussed in \cite{Harr07}.
For self-containedness we briefly explain the approach and introduce the necessary notation.

\subsection{Matrix formulation of the ASEP}

In the context of interacting particle systems with state space
$\mathbb{V}$ the Markov generator $L$ acting on cylinder functions $f(\eta)$ 
of the configuration $\eta \in  \mathbb{V}$ is 
usually defined through the relation
\bel{2-0a}
L f(\eta) = \sum_{\eta'} w_{\eta', \eta} [f(\eta') - f(\eta)].
\ee
where $w_{\eta', \eta}$ is the transition rate from a configuration $\eta$ to $\eta'$.
In particular for a probability measure $\mu(t)$ we have
\bel{2-0b}
\frac{\rmd}{\rmd t} \exval{f}_\mu =\exval{Lf}_\mu
\ee
where $\exval{\cdot}_\mu$ denotes expectation w.r.t.~$\mu$. In particular, taking  $f$ to be
the indicator function ${\bf 1}_\eta$
on a fixed configuration $\eta$ yields the {\it master equation}
\bel{2-1}
\frac{\rmd}{\rmd t} \mu(\eta;t) = 
\sum_{\stackrel{\eta'\in \mathbb{V}}{\eta'\neq\eta}}
\left[w_{\eta, \eta'} \mu(\eta';t) -
w_{\eta', \eta} \mu(\eta;t)\right] 
\ee
for the time evolution of the probability $\mu(\eta;t)$ of finding the configuration 
$\eta$ at time $t$. A stationary distribution, i.e., an invariant measure satisfying
$\rmd \mu/\rmd t=0$ is denoted by $\mu^*$ and a stationary probability of a configuration
$\eta$ is denoted $\mu^*(\eta)$.
%Unless confusion can arise, we shall refer to $\mu(\eta;t)$ simply as probability.

A convenient way to 
write the master equation (\ref{2-1}) in a matrix form is provided by the so-called
quantum Hamiltonian formalism \cite{Lloy96,Schu01}. 
The idea is to assign to each of the possible configurations 
$\eta$ a column vector $\ket{\eta}$ which together 
with the transposed vectors  $\bra{\eta}$ form an orthogonal basis of
a complex vector space with inner product $\inprod{\eta}{\eta'}= \delta_{\eta,\eta'}$.
Here $\delta_{\eta,\eta'}$ is the Kronecker symbol
which is equal to 1 if the two arguments are equal and zero otherwise.
Therefore a measure can be written as a probability vector
\bel{2-3}
| \, \mu(t)\, \rangle = \sum_{\eta \in \mathbb{V}} \mu(\eta;t) \, \ket{\eta}.
\ee
whose components are the probabilities $\mu(\eta;t)= \bra{\eta} \, \mu(t)\, \rangle$.
The bra-ket notation for vectors is an elegant tool borrowed from quantum mechanics.

Using a standard
argument (see for example Chapt. XVII of Feller \cite{feller}), 
for all $t\geq 0$, the master equation (\ref{2-1}) then takes the form of
a Schr\"odinger equation in imaginary time,
\bel{riorita1}
\frac{d}{dt} \ket{\mu(t)} = -H \ket{\mu(t)}
\ee
with the formal solution
\bel{riorita2}
\ket{\mu(t)} = \rme^{-Ht }\ket{\mu(0)}
\ee
reflecting the semi-group property.
The off-diagonal matrix elements $H_{\eta,\eta'}$ of the matrix $H$ are the (negative) transition 
rates $w_{\eta,\eta'}$ and the diagonal entries $H_{\eta,\eta}$
are the sum of all outgoing transition rates $w_{\eta',\eta}$ from 
configuration $\eta$.

For the study of expectations we also define the row vector $\bra{s}$ 
\bel{allvector}
\bra{s} :=(1,1,\ldots, 1)
=\sum_{\eta\in\{0,1\}^{L}}\bra{\eta}
\ee
which we call the summation vector. By conservation of probability we have for any probability
vector $\inprod{s}{\mu(t)}= \sum_{\eta \in \mathbb{V}} \mu(\eta;t) = 1$.
This implies $\bra{s} H = 0$, i.e., the row vector $\bra{s}$
is a left eigenvector of $H$ with eigenvalue 0. A right eigenvector 
with eigenvalue 0 is an invariant measure $\ket{\mu^\ast}$.

Expectation values $\exval{f}=\sum_\eta f(\eta) \mu(\eta)$
of a function $f(\eta)$ are obtained by taking the scalar product
$\bra{s} \hat{f} \ket{\mu}$
of the diagonal matrix
\be
\hat{f} := \sum_{\eta \in  \mathbb{V}} f(\eta) \ket{\eta}\bra{\eta}
\ee
where $\ket{\eta}\bra{\eta}$ is used as a shorthand for $\ket{\eta}\otimes\bra{\eta}$
as is standard in the quantum mechanics literature. 
Moreover, we introduce for $t \geq 0$ the non-diagonal matrices
\bel{Heisenberg}
\hat{f}(t) = \rme^{Ht} \hat{f} \rme^{-Ht}
\ee
and find for any initial measure $\mu(0)$ the useful identity
\be
\exval{f(t)} = \bra{s} \hat{f} \ket{\mu(t)} = \bra{s} \hat{f}(t) \ket{\mu(0}.
\ee
Correspondingly for joint expectations at different times $t_{i+1} \geq t_i \geq 0$
one has
\be
\exval{f_n (t_n) \dots f_2 (t_2) f_1 (t_1)} = \bra{s} \hat{f}_n (t_n) \dots \hat{f}_2 (t_2) \hat{f}_1 (t_1) \ket{\mu(0}.
\ee
Notice that \eref{2-0b} translates into
$\rmd/(\rmd t) \exval{f(t)} = - \exval{\hat{f}H}$.
As there will be no danger of confusing $L$ and $H$ we shall
somewhat loosely refer also to $H$ as generator of the process,
since we can identify $H$ with usual generator acting on
cylinder functions $f$ by its action to the left on $\bra{s} \hat{f} H$.

For the ASEP the vector representation of the state space
$ \mathbb{V}=\{0,1\}^{L}$ is conveniently done using a tensor basis.
The tensor product is denoted by $\ot$, and $A^{\ot k}$ denotes the $k$-fold
tensor product of $A$. The superscript $T$ indicates transposition.
A configuration $\eta\in \{0,1\}^{L}$  will be represented
by the vector  $\ket{\eta}\in \C^{2^{L}}$ which is defined in the
following manner
\bel{represconfig}
\ket{\eta}=\ket{\eta(-M+1)} \ot \ket{\eta(-M+2)} \ot \cdots \ot \ket{\eta(M)} 
\ee
where for each $i=-M+1, \ldots, M$, $\ket{\eta(i)}=(0,1)^T$, if the
$i$-th site in the configuration $\eta$ contains a particle, and
$\ket{\eta(i)}=(1,0)^T$,  otherwise.
Observe that for any configuration $\eta$, the corresponding vector $\ket{\eta}$
has $1$ at one of its
components and $0$ at all others.

In order to construct the matrix representation of the generator $H$ for the
ASEP we introduce the three $2$-by-$2$ matrices
\bel{matrizy}
\sigma^+:=\left(
                \ba{cc} 0 & 1\\
                             0 & 0 
                \ea
\right), \quad
\sigma^-:=\left(
                \ba{cc} 0 & 0\\
                             1 & 0 
                \ea
\right), \quad
\hat{n}:=\left(
                \ba{cc} 0 & 0\\
                             0 & 1 
                \ea
\right), 
\ee
Then, denoting by $\mathds{1}$ the two-dimensional unit matrix, we introduce for each $k \in \T $
\be
\sigma_k^\pm:=\mathds{1}\otimes\cdots \otimes \mathds{1}\otimes 
\sigma^\pm\otimes\mathds{1}\otimes \cdots \otimes \mathds{1},
\quad \hat{n}_k:=\mathds{1}\otimes\cdots \otimes \mathds{1}\otimes n\otimes\mathds{1}
\otimes \cdots \otimes \mathds{1}
\ee
where a non-identity matrix appears at the $(M+k)$-th position, counting
from the left to the right.
We also introduce $\hat{v}:=\mathds{1}-\hat{n}$ and $\hat{v}_k = 
\mathds{1}^{\otimes L} - \hat{n}_k$.

For readers unfamiliar with this construction we point out some simple properties of 
these matrices and their action on vectors. Notice that $(1,0)\sigma^+=(0,1)\hat{n}=(0,1)$,
$(0,1)\sigma^-=(1,0)$, $(1,0)\hat{n}=(1,0)\sigma^-=(0,1)\sigma^+=(0,0)$; and analogously,
$\sigma^+(0,1)^T=(1,0)^T$, $\sigma^-(1,0)^T=\hat{n}(0,1)^T=(0,1)^T$,
$\hat{n}(1,0)^T=\sigma^+(1,0)^T=\sigma^-(0,1)^T=(0,0)$. These identities and  the representation
\eref{represconfig} yield the following properties of $\sigma^\pm_k$ and $\hat{n}_k$:
If a configuration $\eta$ does not have a particle at the site $k$ then
$\ket{\eta^\prime}:=\sigma^-_k\ket{\eta}$ corresponds to the configuration that
coincides with $\eta$ on $\T_L\setminus\{k\}$ and has a particle at $k$; if to
the contrary $\eta$ has a particle at $k$ then $\sigma^-_k\ket{\eta}=0$, i.e., 
the vector with all components equal to $0$. If a configuration $\eta$ has a particle at the site $k$ then
$\ket{\eta^\ast}:=\sigma^+_k \ket{\eta}$ corresponds to the configuration that
coincides with $\eta$ on $\T_L\setminus\{k\}$ and does not have  a particle at
$k$; if to the contrary $\eta$ does not have a particle at $k$ then
$\sigma^+_k\ket{\eta}=0$.
Using the equality $(\sigma^+_k)^T=\sigma^-_k$ and the above notations, we have that
if $\eta$ does not have a particle at $k$ then $\bra{\eta}\sigma^+_k=\bra{\eta^\prime}$ 
and $\bra{\eta}\sigma^-_k=0$, while if $\eta$ has a
particle at $k$ then $\bra{\eta}\sigma^+_k=0$ and $\bra{\eta}\sigma^-_k=\bra{\eta^\ast}$.
Accordingly, $\sigma^-_k$ and $\sigma^+_k$ are called {\it the
particle creation/annihilation operators}.
The operator $\hat{n}_k$ is called {\it the number operator};
when applied to $\ket{\eta}$, it returns $\ket{\eta}$, if there was a particle
at the site $k$, and results in $0$ otherwise, i.e. we have
\bea
\hat{n}_k \ket{\eta} & = & \eta(k) \ket{\eta} = 
\left\{ \ba{ll} \ket{\eta} & \mbox{if } \eta(k) = 1 \\ 0 & \mbox{if } \eta(k) = 0 \ea \right. , \\
\hat{v}_k \ket{\eta} & = & (1-\eta(k)) \ket{\eta} =
\left\{ \ba{ll} \ket{\eta} & \mbox{if } \eta(k) = 0 \\ 0 & \mbox{if } \eta(k) = 1 \ea \right. .
\eea

With these definitions the generator of the ASEP becomes
\bel{gamiltonian}
H:=-\sum_{i=-M+1}^{M} \left[
c_r (\sigma^+_i \sigma^-_{i+1} - \hat{n}_i \hat{v}_{i+1})+
c_\ell (\sigma^-_i \sigma^+_{i+1} - \hat{v}_{i} \hat{n}_{i+1} )\right].
\ee
For an arbitrary distribution $\mu(0)$ on $\{0,1\}^{L}$ and its vector representation
$\ket{\mu(0)}$ as defined above we denote
by $\ket{\mu(t)}$ the vector representation of the  distribution
of ASEP at time $t$, starting from $\mu(0)$. 
We point out that in the tensor basis a product
measure is represented by a tensor product of the single-site marginals.
For the summation vector we have the tensor representation
$\bra{s} =(1,1)^{\otimes L}$.

\subsection{Grandcanonical conditioning on an atypical current}

The generator $H$ defined above is for the hopping dynamics only, it does not 
include the evolution of the integrated current $J(t)$. Since $J(t)$ can take any integer value $J$,
the state space for the full process is $\{0,1\}^L\times \Z$. 
In order to construct the matrix representation we choose as basis for this space
the set of product vectors $\ket{\eta}\otimes\ket{J}$ with $J\in \Z$. 
(Notice that the symbols $\bra{\cdot}$ and $\ket{\cdot}$ can be vectors in different vector
spaces, as will be clear from the form the argument.) Defining (infinite-dimensional)
operators $A^\pm$ through the relation $A^\pm \ket{J} = \ket{J\pm 1}$
we obtain following Ref. \cite{Harr07} the generator
\bel{fullmatrix}
G:=-\sum_{i=-M+1}^{M} \left[
c_r(\sigma^+_i\sigma^-_{i+1}\otimes A^+-n_i v_{i+1})+
c_\ell(\sigma^-_i\sigma^+_{i+1}\otimes A^- -v_{i} n_{i+1}) \right]
\ee
for the full process. One sees
that the elementary hopping matrices not only change the configuration $\eta$
according to which jump has occurred, but also change the value of total integrated 
current $J$ accordingly. An initial configuration is represented by a tensor vector
$\ket{\eta,J} := \ket{\eta}\otimes\ket{J}$. We shall take $J(0)=0$ as initial
value of $J(t)$. Hence an initial measure for the full process has the form
$\ket{\mu(0)}\otimes\ket{0}$. 
In order to compute expectation values we define the extended
summation vector $\bra{\hat{s}} := \bra{s}\otimes\bra{s'}$ where $\bra{s'}$ is the
infinite-dimensional summation vector for the counting variable $J$.

Consider now the process conditioned on $J(t)$ 
having reached some fixed value $J$ at time $t\geq 0$.
The conditional distribution $\mu(\eta,J;t)$ of the ASEP under this conditioning has the form
\be
\mu(\eta,J;t) = \frac{\bra{\eta,J} \rme^{-Gt} \ket{\mu(0),0}}{ Z_J(t)}
\ee
where 
\be
Z_J(t) := \Prob{J(t)=J} = \bra{s,J} \rme^{-Gt} \ket{\mu(0),0}
\ee
is the marginal distribution of the integrated current and $\bra{s,J}:=\bra{s}\otimes\bra{J}$ with
$J\in \Z$.
We write the conditional expection of some function $f$ of the occupation numbers of the ASEP
as
\bel{3-1}
\exval{f(t)}_J = \frac{ \exval{f;J(t)=J}}{Z_J(t)} = \frac{\bra{s,J} \hat{f} \rme^{-Gt} \ket{\mu(0),0}}{Z_J(t)}.
\ee

In actual fact, however, we are not interested in conditioning on a fixed value $J$ of the current,
but rather in expectations for an ensemble where the current is allowed to fluctuate around some
generally atypical value.  To this end we follow standard procedure and define a generalized
fugacity $y = \rme^s$ and a ``grand canonical'' current
ensemble
\be
Y_s(t) =  \exval{y^{J(t)}}= \sum_{J \in \Z} y^J Z_J(t).
\ee
Analogously we define the fluctuating conditional probability of a configuration $\eta$ as
\be
\mu_s(\eta;t) = \frac{\sum_{J \in \Z} y^J \bra{\eta,J} \rme^{-Gt} \ket{\mu(0),0}}{ Y_s(t)}
\ee
and the corresponding
expectations
\bel{EBP3-2}
\exval{f(t)}_s  := \frac{\exval{f(t)y^{J(t)}}}{ \exval{y^{J(t)}}}
= \frac{\sum_{J \in \Z} y^J \exval{f;J(t)=J}}{Y_s(t)}.
\ee
Notice that $Y_0(t) = 1$ and that therefore $\ket{\mu_0} = \ket{\mu}$
is the usual (unconditioned) measure and $\exval{f(t)}_0$ is the usual
(unconditioned) expectation of $f$. We shall refer to the latter quantity
as the ``typical'' expectation. For $s\neq 0$ we call corresponding quantities
atypical.

Consider now the
unnormalized fluctuating conditional probability
\be
\tilde{\mu}_s(\eta;t) = \sum_{J \in \Z} y^J \bra{\eta,J} \rme^{-Gt} \ket{\mu(0),0}.
\ee
The following simple result is often used in the literature without proof, see e.g. \cite{Derr98} 
and, for a formal 
derivation using counting operators $A^\pm$, see \cite{Harr07}. 
One has for the associated unnormalized probability vector
\be
\ket{\tilde{\mu}_s(t)}= \rme^{-\tilde{H}(s)t} \ket{\mu(0)}
\ee
and
\be
Y_s(t) = \bra{s} \rme^{-\tilde{H}(s)t} \ket{\mu(0)}
\ee
for the normalization.
Here $\tilde{H}(s)$ is the matrix of dimension $2^L$ obtained from $G$ by substituting the 
counting operators $A^\pm$ by the $c$-numbers $\rme^{\pm s}$. Notice that 
$\tilde{H}(0)=H$ is the generator of the ASEP constructed above. For $s\neq 0$ the matrix $\tilde{H}(s)$
does not conserve probability, but nevertheless has an intuitive probabilistic interpretation.
It gives a weight $\rme^{\pm s}$ to each transition in a particular realization of the process.
Hence we shall refer to $\tilde{H}(s)$ as the weighted generator. The generalized chemical potential
$s$ parametrizes the mean current of this weighted process. 
Thus we arrive at the central object of our interest, which is the conditional distribution 
\bel{conddist}
\ket{\mu_s(t)} := \frac{ \rme^{-\tilde{H}(s) t} \ket{\mu(0)}}
{\bra{s} \rme^{-\tilde{H}(s) t} \ket{\mu(0)}}.
\ee
This quantity describes the approach to the long-time large deviation regime from a given initial
distribution and hence provides information about the space-time structure of the long-time large deviation 
regime. Notice that below we shall drop the subscript $s$ which indicates the
$s$-dependence.
(Grandcanonical) conditioned expectations
at time $t$
are then computed as follows:
\bel{condexp}
\exval{f(t)}_s
= \frac{\bra{s} \hat{f} \rme^{-\tilde{H}(s) t} \ket{\mu(0)}}
{\bra{s} \rme^{-\tilde{H}(s) t} \ket{\mu(0)}} .
\ee

In the same spirit one can investigate the space time structure of the process under the condition that the
integrated current across some fixed bond $(k,k+1)$ has attained a certain value. Going to the grandcanonical
conditioning leads to a matrix $\tilde{H}^{(k)}(s)$ where only the hopping terms for the bond $(k,k+1)$ have the
weights $\rme^{\pm s}$. As discussed in the introduction we refer to this setting as 
{\it local conditioning}, as opposed to the {\it global conditioning} involving the global current. Notice that the 
global time-integrated current $J(t)$ is (trivially)
extensive in system size $L$ and hence its generating function
does not have a good limit for $L\to\infty$. Below we shall choose
under global conditioning the coefficient $s$ to be order $1/L$ as this yields the generating function for the current 
density $j$. This quantity has finite expectation $\exval{j}$ even for $L\to \infty$.

\section{Shock measures}

We shall consider the evolution of two distinct types of shock measures. 
First we recall the definition of a shock measure with shock at site $m$ in the
infinite integer lattice $\Z$ as a Bernoulli product measure with marginal fugacity $z_1$ up to
site $m-1$ and fugacity $z_2 > z_1$ from site $m$ onwards \cite{Beli02}.
We denote these measures by  $\mu^+_m$. Likewise one can define an antishock 
measure $\mu^-_n$ for the infinite lattice with antishock
at site $n$, i.e., fugacities $z(k)=z_2$ for $k \leq n$ and $z(k)=z_1$ for $k > n$.

Type I shock measures for the torus $\T_L$
are defined in analogy to these shock measures as follows: 
\begin{df}
\label{shock1}
For the torus $\T_L$ a shock measure $\mu^{I}_{m,n}$ of type I with $n\neq m$ is a 
Bernoulli product measure with fugacities 
$$\left.
\ba{ll} z_2 & \mbox{at the set of sites } P^{m<n}_{high} := \{ k \in \T_L : m < k \leq n \}\\
z_1 & \mbox{at the set of sites } P^{m<n}_{low} := \T_L \setminus P_h \ea
\right\} \mbox{\rm  if } -M+1 \leq m < n \leq M
$$
and
$$\left.
\ba{ll} 
z_1 & \mbox{at the set of sites } P^{n<m}_{low} := \{ k \in \T_L : n < k \leq m \} \\
z_2 & \mbox{at the set of sites } P^{n<m}_{high} := \T_L \setminus P_l
\ea
\right\} \mbox{\rm  if } -M+1 \leq n < m \leq M.
$$
\end{df}
Due to Condition S and the convention $c_r > c_\ell$ we have $z_2>z_1$ and therefore
$\rho_2 > \rho_1$. Hence in the definition
\eref{shock1} the first index marks the site after which the high density region $P_{high}$ begins, 
which extends up to
site $n$, counted modulo $L$ in the principal domain  $\{-M+1, \dots, M\}$. 
We call site $m$ the {\it microscopic position of the shock} and site $n$ the
{\it microscopic position of the antishock} in the shock measure of type I. 

In vector representation we have
\bel{vecshock1a}
\ket{\mu^I_{m,n}} =  \left\{ \ba{ll} \frac{1}{A^+_{m,n}}
\left(\ba{c} 1 \\ z_1 \ea \right)^{\otimes (m+M)} 
\otimes \left(\ba{c} 1 \\ z_2 \ea \right)^{\otimes (n-m)}
\otimes \left(\ba{c} 1 \\ z_1 \ea \right)^{\otimes (M-n)} & m < n \\
\frac{1}{A^-_{m,n}} \left(\ba{c} 1 \\ z_2 \ea \right)^{\otimes (n+M)} 
\otimes \left(\ba{c} 1 \\ z_1 \ea \right)^{\otimes (m-n)}
\otimes \left(\ba{c} 1 \\ z_2 \ea \right)^{\otimes (M-m)} &  n < m
\ea \right.
\ee
with $A^+_{m,n} = (1+z_1)^{2M+m-n}(1+z_2)^{n-m}$ and
$A^-_{m,n} = (1+z_2)^{2M-m+n}(1+z_1)^{m-n}$.
Tensor products with exponent 0 are defined to be absent.
We remark that in terms of densities $\rho_i = z_i/(1+z_i)$ these vectors read
\bel{vecshock1}
\ket{\mu^I_{m,n}} = \left\{ \ba{ll}
\left(\ba{c} 1-\rho_1 \\ \rho_1 \ea \right)^{\otimes (m+M)} 
\otimes \left(\ba{c} 1-\rho_2 \\ \rho_2 \ea \right)^{\otimes (n-m)}
\otimes \left(\ba{c} 1-\rho_1 \\ \rho_1 \ea \right)^{\otimes (M-n)} & m < n\\
\left(\ba{c} 1-\rho_2 \\ \rho_2 \ea \right)^{\otimes (n+M)} 
\otimes \left(\ba{c} 1-\rho_1 \\ \rho_1 \ea \right)^{\otimes (m-n)}
\otimes \left(\ba{c} 1-\rho_2 \\ \rho_2 \ea \right)^{\otimes (M-m)} &  n < m.
\ea \right.
\ee
The unnormalized (!) restriction of $\mu^{I}_{m,n}$ to the sector with $N$ particles is 
denoted by $\mu^{I,N}_{m,n}$, i.e.,
\bel{projection}
\ket{\mu^{I,N}_{m,n}} = P_N \ket{\mu^{I}_{m,n}}
\ee
where the projector on configurations with $N$ particles
is defined by
\bel{projector}
P_N \ket{\eta} = \left\{ \ba{ll} 
\ket{\eta} & \mbox{if } \sum_{k\in\T_L} \eta(k) = N \\
0 & \mbox{otherwise }
\ea \right ..
\ee
Shock measures of type I 
are a particular microscopic realization of a shock/antishock pair in a macroscopic step function 
density profile 
with density $\rho_2$ in the interval $[x,y)$ of rescaled coordinates $m \to x$, $n \to y$ (modulo 1) under 
suitable rescaling of space. 
For $M \to \infty$, 
$n \to \infty$ and $m$ fixed we recover the Bernoulli shock measures $\mu^+_m$ of \cite{Beli02}
for the ASEP defined on $\Z$ fugacities $z(k)=z_1$ for $k \leq m$ and $z(k)=z_2$ for $k > m$.
In a similar fashion one can recover an antishock measure $\mu^-_n$ by taking the thermodynamical
limit such that $n$ remains fixed and both $M$ and $m$ are taken to infinity.

For $n=M$ we shall drop the second subscript and write $\mu^{I}_{m} := \mu^{I}_{m,M}$
and similarly for the vectors and the projections on $N$ particles.
For $n=M$ and $m=\pm M$  the shock measures reduce to the usual Bernoulli
product measures which we denote by 
\bel{Bernoulli}
\ket{\mu_{y}} =  \frac{1}{(1+y)^L}
\left(\ba{c} 1 \\ y \ea \right)^{\otimes L} = \left(\ba{c} 1-\rho \\ \rho \ea \right)^{\otimes L} 
\ee
 where $y = \rho/(1-\rho) \in [0,\infty)$ is the fugacity. In particular,
\be
\label{Bernoulli12}
\ket{\mu^I_{M,M}} \equiv \ket{\mu^I_{M}} = \ket{\mu_{z_1}} , 
\quad \ket{\mu^I_{-M,M}}  \equiv \ket{\mu^I_{-M}} = \ket{\mu_{z_2}} .
\ee
We stress that $\mu^I_M \neq \mu^{I}_{-M}$.

We shall make use of the transformation property
\begin{lmm}
\label{Bernoullitrafo}
Let $\ket{\mu_{y}}$ be the vector representation of the Bernoulli product measure with fugacity $y$
for $L$ sites and $\hat{N}_m = \sum_{k=m+1}^M \hat{n}_k$ be the partial number operator. 
Then $\forall  z_1, \, z_2 \in (0,\infty)$ and $-M\leq m \leq M$
\bel{Bernoullishocktrafo}
\ket{\mu^I_{m}} = \left( \frac{1+z_1}{1+z_2} \right)^{M-m} \left( \frac{z_2}{z_1} \right)^{\hat{N}_m} \ket{\mu^I_{M,M}},
\ee
and for fixed particle number $N$
\bel{BernoullisN}
\ket{\mu^{I,N}_{-M}} = \left( \frac{1+z_1}{1+z_2} \right)^L \left( \frac{z_2}{z_1} \right)^{N} \ket{\mu^{I,N}_{M,M}}.
\ee
\end{lmm}
{\it Proof:}
With the matrix representation of the number operator $\hat{n}$ for a single site
$$y^{\hat{n}} = \mathds{1} + (y-1) \hat{n} = \left( \ba{cc} 1 & 0 \\ 0 & y \ea \right),$$
the tensor property $y^{\hat{N}_m} = \mathds{1}^{\otimes M+m} \otimes (y^{\hat{n}})^{\otimes M-m}$
and the vector representation \eref{vecshock1a} the proof of the first equality becomes elementary
multilinear algebra. The second equality
follows from $\hat{N}_{-M} = \hat{N}$ and the fact that the projection on $N$ sites can be interchanged 
with the number operator
and that the projected vector is an eigenstate with eigenvalue $N$ of the number operator $\hat{N}$.
\qed

Notice that when we work with condition S, the family of shock measures of type I is
a one-parameter family of measures indexed by $z_1$. In particular, \eref{Bernoullishocktrafo}
may be written
\bel{Bernoullishocktrafo2}
\ket{\mu^I_{m}} = \left( \frac{1+z_1}{1+q^2 z_1} \right)^{M-m}  q^{2\hat{N}_m} \ket{\mu_{z_1}}
\ee
and for $m=-M$ we have the analogue of \eref{BernoullisN} for the sector of fixed $N$:
\bel{BernoullisN2}
\ket{\mu^{I,N}_{-M}} = \left( \frac{1+z_1}{1+q^2 z_1} \right)^{L}  q^{2N} \ket{\mu^N_{z_1}}.
\ee

Next we introduce a new one-parameter family of shock measures which -- to our knowledge -- has 
not yet been 
considered in the literature. 

\begin{df}
\label{shock2}
For $ -M+1 \leq m \leq M-1$ a shock 
measure $\mu^{II}_{m}$ of type II  on $\T_L$ with shock at position $m$
is a Bernoulli product measure
with space-dependent fugacities
\be
\label{shock2alt1a}
z(k) = \left\{ \ba{ll} 
z_1 q^{2\frac{m-M-k}{L}} & \mbox{\rm for } - M < k \leq m \\
z_1 q^{2\frac{m+M-k}{L}} & \mbox{\rm for } m < k \leq M
\ea \right\}  .
\ee
Furthermore, $\mu^{II}_{-M}\equiv \mu^{II}_{M}$ is a Bernoulli product measure
with space-dependent fugacities
\be
\label{shock2alt1b}
z(k) = 
z_1 q^{\frac{-2k}{L}}  \mbox{\rm for } - M < k \leq M.
\ee
\end{df}

We write down the vector presentation of these measures
\bel{vecshock2a}
\ket{\mu^{II}_{m}} = \frac{1}{Z_{m}}
\left(\ba{c} 1 \\ z({-M+1}) \ea \right) \otimes 
\left(\ba{c} 1 \\ z({-M+2}) \ea \right) \otimes \dots
\otimes \left(\ba{c} 1 \\ z(M) \ea \right) 
\ee
with
\be
\label{Z+}
Z_{m} = \prod_{k=-M+1}^m \left(1+z_1q^{2\frac{m-M-k}{L}}\right) 
\prod_{k=m+1}^M \left(1+z_1q^{2\frac{m+M-k}{L}}\right) \quad (m<n).
\ee
The unnormalized restrictions of these measures on the sectors with $N$ particles 
are denoted $\mu^{II,N}_{m}$.

Shock measures of type I and type II are related by a similarity transformation.
This is the content of the following lemma. 
\begin{lmm}
\label{Lemma2}
Assume condition S and  let 
\bel{U}
U := q^{-\frac{2}{L} \sum_{k\in \T_L} k \hat{n}_k}.
\ee
Then for $-M \leq m  \leq M$ we have
\bel{trafo}
\ket{\mu^{II}_{m}} = \frac{1}{Y_{m}} q^{\frac{2(m-M)}{L}\hat{N}}  U \ket{\mu^{I}_{m}}
\ee
with
\bel{Ym}
Y_{m} = \prod_{k=-M+1}^m \left(\frac{1+z_1q^{2\frac{m-M-k}{L}}}{1+z_1}  \right) 
\prod_{k=m+1}^M \left(\frac{1+z_1q^{2\frac{m+M-k}{L}}}{1+q^2z_1}\right).
\ee
Moreover, for fixed particle number $N$
\be
\ket{\mu^{II,N}_{m}} = \frac{1}{Y_{m}} q^{\frac{2(m-M)}{L}N}  U \ket{\mu^{I,N}_{m}}.
\ee

\end{lmm}

{\it Proof:} Observing that $z_1 q^{2\frac{2M}{L}} = z_1 q^2 = z_2$
the proof works like for lemma \eref{Bernoullitrafo}, but using also that not only 
the projector $P_N$ on $N$ particles and the number operator $\hat{N}$ 
are diagonal in the basis used 
throughout this work, but also the transformation $U$.
Hence all three operations can be arbitrarily interchanged.
 \qed

\begin{figure}
\centering
\includegraphics[width=0.6\textwidth]{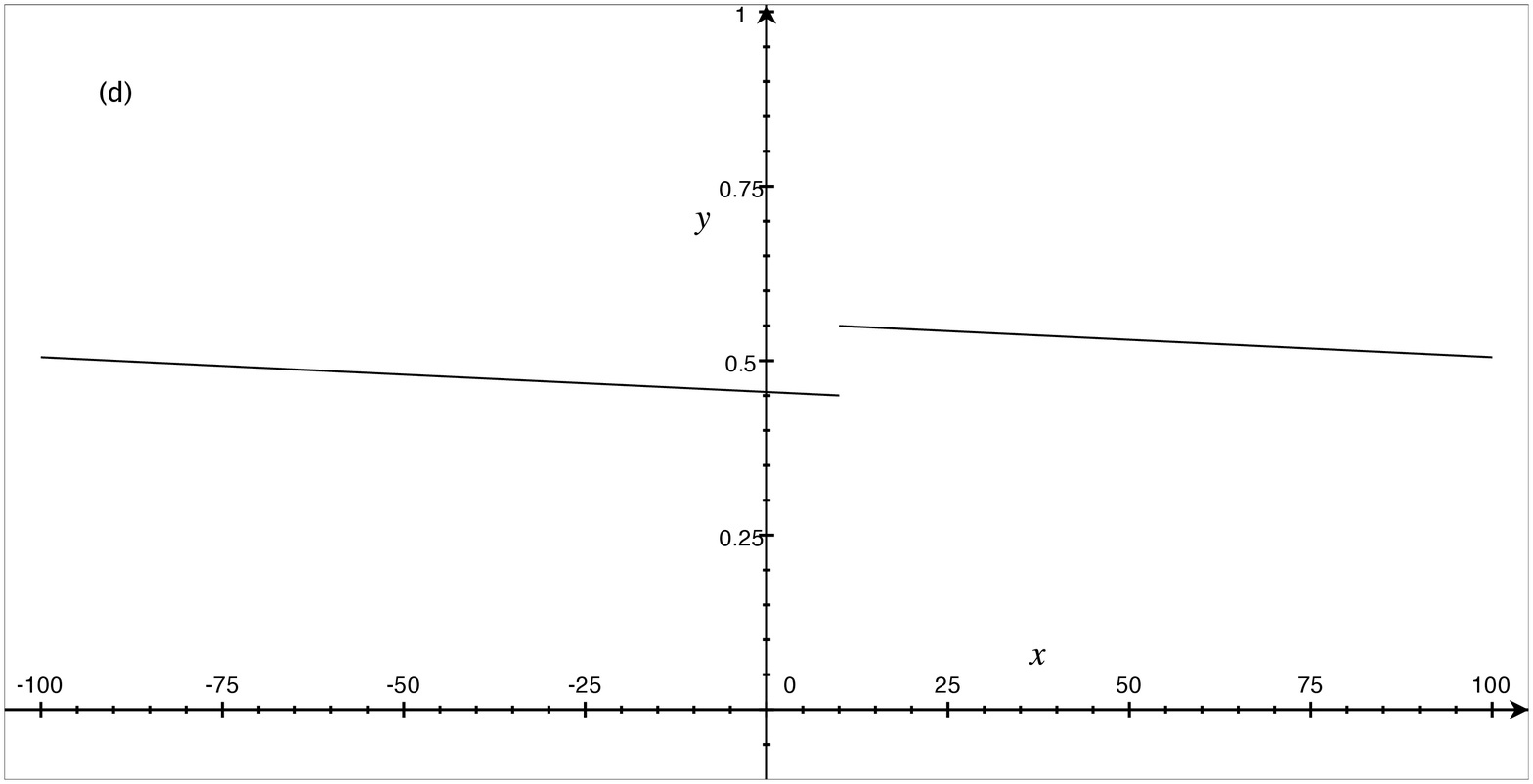}\vspace*{5mm}
\includegraphics[width=0.6\textwidth]{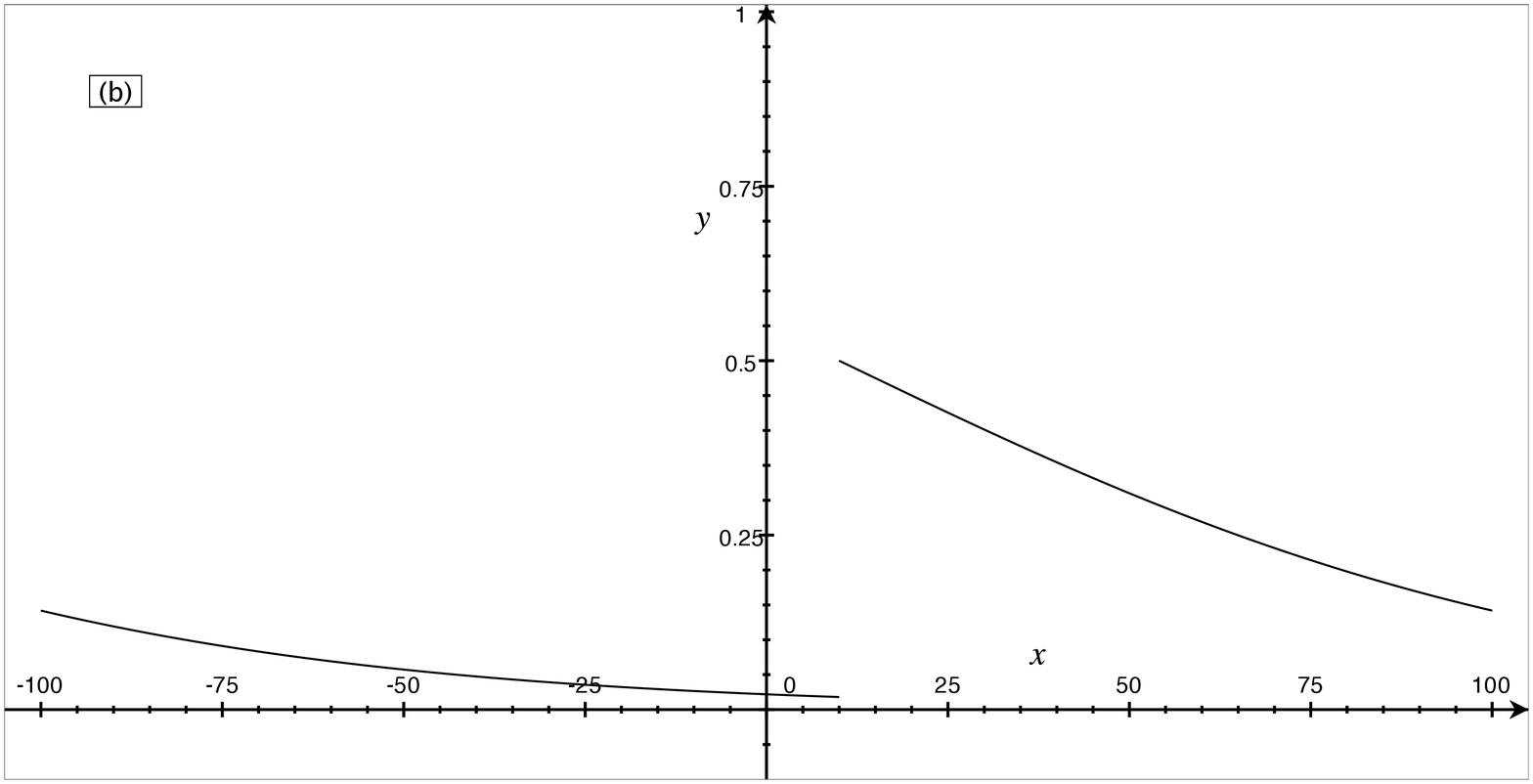}\vspace*{5mm}
\includegraphics[width=0.6\textwidth]{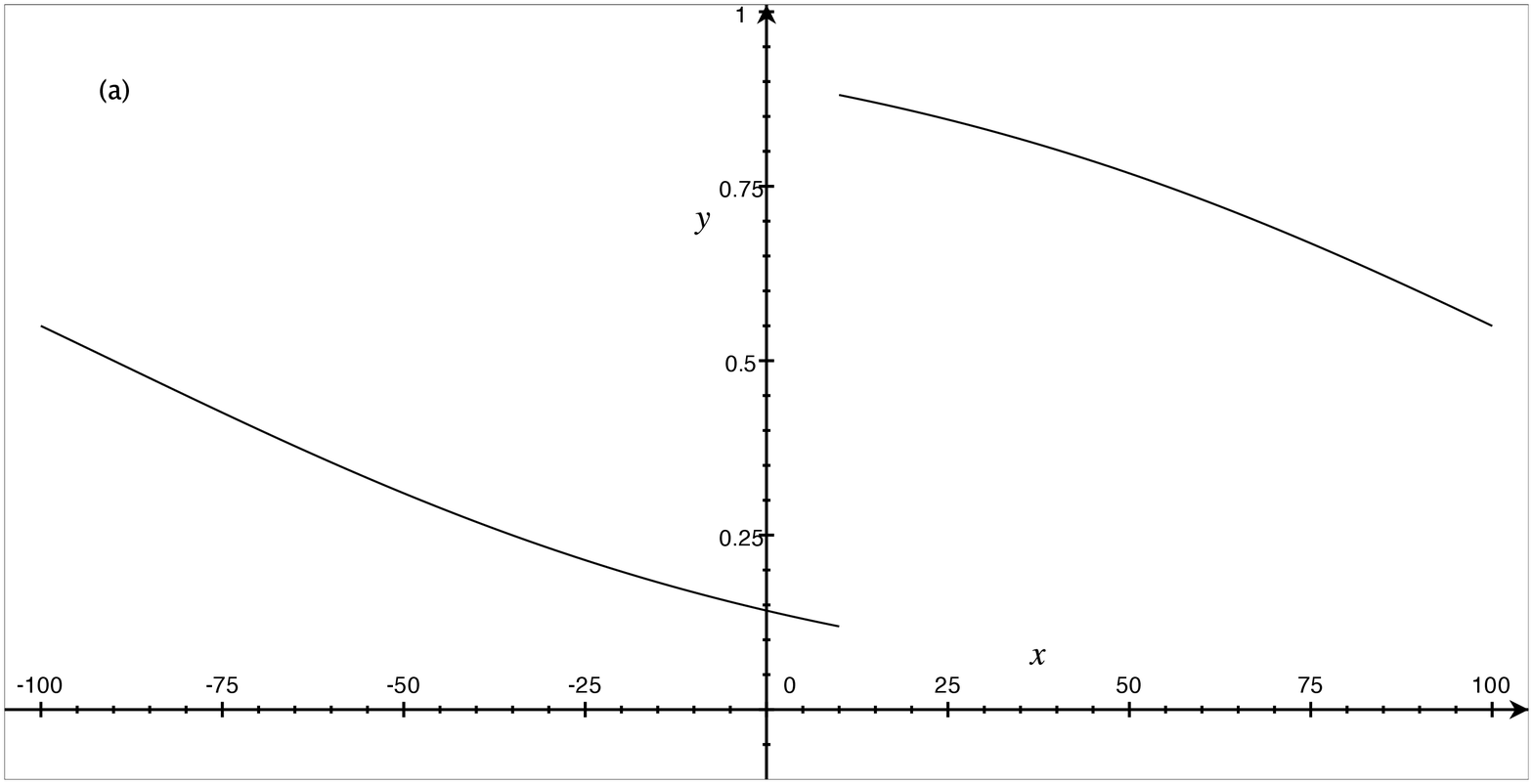} \vspace*{5mm}
\includegraphics[width=0.6\textwidth]{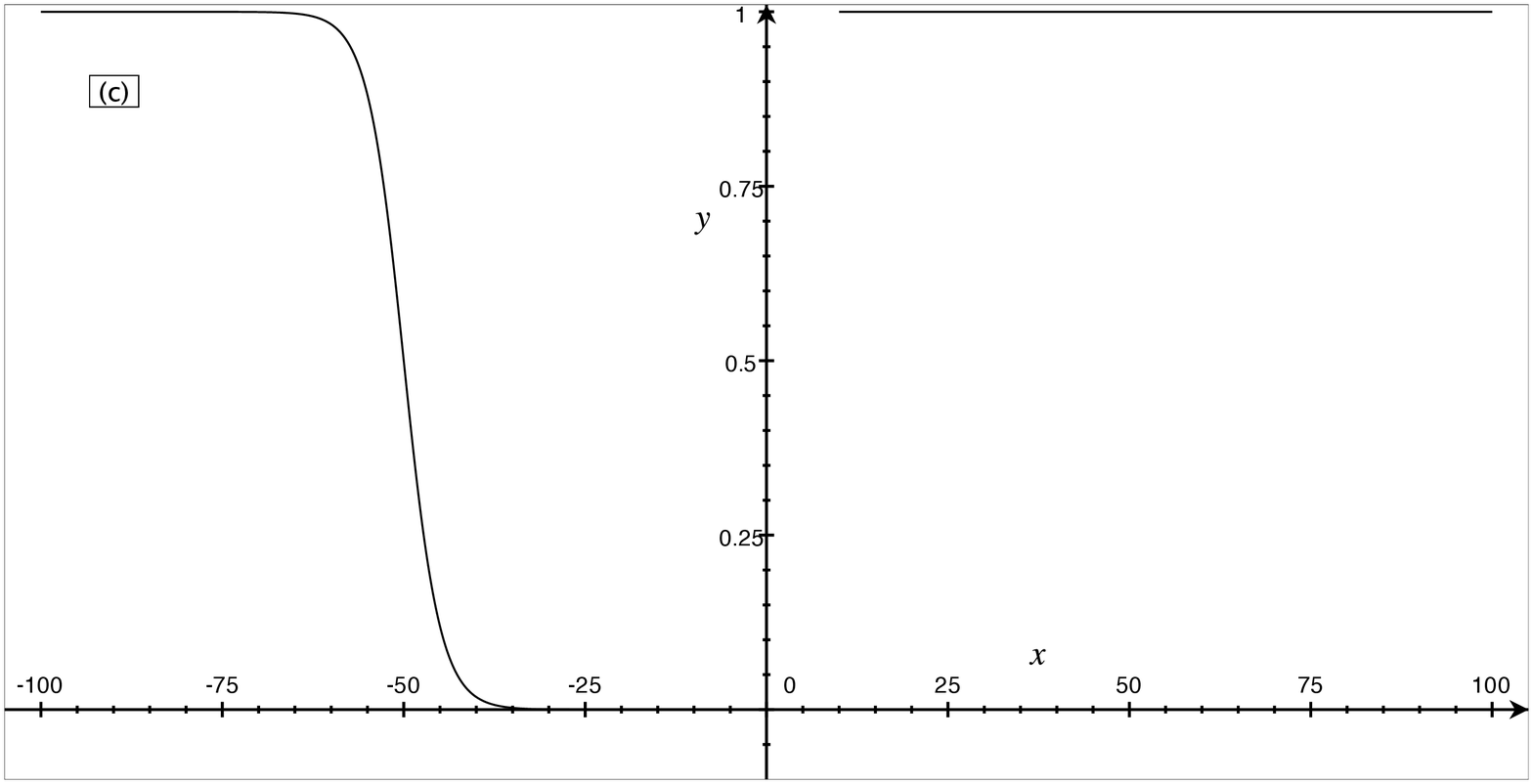}
\caption{Interpolated density profiles for type II shock measures for lattice size $L$=200, shock position 
$m=10$ and different asymmetries and total densities: (a) $E=0.2$, $c=0.5$,  (b) $E=2$, $c=1$, (c)
$E=2$, $c=0.5$,  (d) $E=40$, $c=0.3$.}
\label{typeIIshockprofiles}
\end{figure}

To elucidate the nature of these type II shock measures we  define parameters
$c$, $E$ through $\rme^{2E} := q^2$, $\rme^{- 2E c} := z_1$. Using Condition S gives
$z_2 = \rme^{- 2E (c-1)}$ and we can then express
$\mu^{II}_{m}$ as a Bernoulli product measure with densities
\be
\rho(k) = \left\{ \ba{ll}
\frac{1}{2} \left[1 - \tanh{\left(\frac{E(k+cL)}{L}\right)}\right] & -M < k \leq m \\
\frac{1}{2} \left[1 - \tanh{\left(\frac{E(k+(c-1)L)}{L}\right)}\right] & m < k \leq M
\ea \right. .
\ee
For hopping asymmetry $q$ of order 1 the variation of density between neighbouring sites is of order 
$1/L$, except between site $m$ and $m+1$ where there is a jump of order 1. On the other hand, for strong
asymmetry with $E$ of order $L$ the density jumps from nearly 0 to nearly 1 at the shock position $m$ and there
is a sharp transition from nearly 1 to nearly 0, extending over a few lattice sites in the vicinity of 
$k^\ast=-cL$ (modulo $L$). Here "nearly" means up to corrections exponentially small in $L$. 
This sharp transition is the microscopic analogue of an antishock in the type II shock measure
which exists only for strong asymmetry. For small
asymmetry the 
density profile is almost linear, see \Fref{typeIIshockprofiles}.

\section{Microscopic conditioned evolution of shocks}

\subsection{Main results}

We are now in a position to state and prove our main results on the time
evolution of the ASEP
under conditioned dynamics, viz. for type II shocks
under global conditioning, and for type I shocks under local conditioning.
Throughout this section we assume condition S to be satisfied.

\begin{theo}
\label{main} (Shock evolution under global conditioning)
Consider the ASEP on $\T_L$ with $N$ particles and hopping rates
$c_\ell$ and $c_r$ to the left and to the right respectively.
Let the initial distribution at time $t=0$ be given
by the type II shock measure $\mu^{II,N}_{m}$ and let $\mu^{II,N}_{m}(t)$ denote the grandcanonically conditioned
distribution of the ASEP at time $t$ under global conditioning with weight $\rme^{sL} = c_\ell/c_r$. Then
for $m \in\T_L $ and any $t\geq 0$
\bel{Theorem1b}
 \mu^{II,N}_{m}(t) = \sum_{l=-M+1}^{M}p_t(l \iz m) \mu^{II,N}_{l}
\ee
where 
$$
p_t(l\iz m)= e^{-(\tilde{c}_r + \tilde{c}_\ell)t}
 \sum_{p=-\infty}^\infty
\left( \frac{\tilde{c}_r}{\tilde{c}_\ell} \right)^{(l-m+pL)/2}
I_{m-l+pL}(2\sqrt{\tilde{c}_r \tilde{c}_\ell}\, t)
$$ 
with the modified Bessel function $I_{n}(\cdot)$ and $\tilde{c}_r = c_r q^{-\frac{2N}{L}}$ and
$\tilde{c}_\ell = c_\ell q^{\frac{2N}{L}}$. 
\end{theo}

We remark that $p_t(l\iz m)$ is the probability that a particle that performs a
continuous-time simple random walk on $\T_L$ with hopping rates
$\tilde{c}_r$ to the right and $\tilde{c}_\ell$ to the left, respectively, 
is at site $l$ at time $t$,
starting from site $m$. Hence
the interpretation of \eref{Theorem1b} is that at any time $t\geq 0$ the distribution 
of this process is a convex combination of shock measures of a 
{\it conserved microscopic
structure
equal to that of the initial one}. The probabilistic interpretation of the 
weights in this convex combinations allows us to say that the shocks (or 
better, the shock positions) perform continuous time random walks on $\T_L$ 
with hopping rates $\tilde{c}_r$ (to the right) and $\tilde{c}_\ell$
(to the left). Seen from the shock position the measure is invariant. 
As opposed to the macroscopic
results of \cite{Bodi05} our results provide exact information for any
$t\geq 0$ and for finite lattice distances. Notice however, that our results require the
choice $\rme^{sL} = c_\ell/c_r$ with arbitrary asymmetry $q$, while
the phenomena observed in \cite{Bodi05} are valid for any $s$, but require weak asymmetry. 

\noindent
\textit{Proof of Theorem \eref{main}.}
In order to prove \eref{Theorem1b}  we observe that with \eref{Bernoullishocktrafo2} (which follows
from \eref{Bernoulli12} and \eref{Bernoullishocktrafo} in Lemma \eref{Bernoullitrafo}) and
with Lemma \eref{Lemma2} we can write the 
initial measure
\bel{pr1}
\ket{\mu^{II,N}_{m}} = V_m \ket{\mu^{N}_{z_1}}
\ee
with
\bel{pr2}
V_m =  \frac{1}{W} q^{\frac{2}{L} \left(\sum_{k=-M+1}^m (-M+m-k) \hat{n}_k + 
\sum_{k=m+1}^M (M+m-k) \hat{n}_k \right)}
\ee
and a non-zero normalization factor
\be
W = (1+z_1)^L \prod_{k=-M+1}^M \left(1+z_1 q^{-\frac{2k}{L}}\right)
\ee
that does not depend on $m$. 
We recall \eref{riorita1} and get for the unnormalized measure
\bel{pr4}
\frac{\rmd}{\rmd t} \ket{\tilde{\mu}^{II,N}_{m}(t)} =  - V_m \tilde{H}^{(m)} \ket{\tilde{\mu}^{N}_{z_1}(t)}
\ee
with $\tilde{H}^{(m)} = V_m^{-1} \tilde{H} V_m$. 

Similarity transformations of the diagonal number operators on the particle creation and annihilation operators 
are of the form
\bel{similar}
\rme^{-a_k \hat{n}_k} \sigma_l^\pm \rme^{a_k \hat{n}_k} = 
\left\{ \ba{ll} \sigma_l^\pm & k\neq l \\
\rme^{\pm a_l} \sigma_l^\pm & k = l \ea \right. .
\ee
From \eref{similar} and Condition S we thus find
\bea
\label{pr5a}
\tilde{H}^{(m)} & = & - \sum_{i=-M+1}^{M}\hspace*{-4mm}^\prime  
\left[ c_r (\sigma^+_i \sigma^-_{i+1} - \hat{n}_i \hat{v}_{i+1})+
c_\ell (\sigma^-_i \sigma^+_{i+1} - \hat{v}_{i} \hat{n}_{i+1} )\right] \\
\label{pr5b}
& & -  \left[
c_\ell \sigma^+_m \sigma^-_{m+1} - c_r \hat{n}_m \hat{v}_{m+1}+
c_r \sigma^-_m \sigma^+_{m+1} - c_\ell \hat{v}_{m} \hat{n}_{m+1} \right]
\eea
where the prime at the summation indicates the absence of  the term with $i=m$.
%We write this in the form $\tilde{H}^{(m)} = \tilde{H}^{(m)}_0 + B^{(m)}_0$ with bulk term $H_m$ given by
%\eref{pr5a} boundary term $B_m$ given by \eref{pr5b}.
We add $-(c_r - c_\ell) (\hat{n}_{i+1} - \hat{n}_i)$ to each term in \eref{pr5a} and 
$-(c_r - c_\ell) (\hat{n}_{m+1} - \hat{n}_m)$
to \eref{pr5b}.
Since the sum of these terms is zero we arrive at 
\bea
\label{pr6a}
\tilde{H}^{(m)} & = & - \sum_{i=-M+1}^{M}\hspace*{-4mm}^\prime  \left[
c_r (\sigma^+_i \sigma^-_{i+1}  - \hat{v}_{i} \hat{n}_{i+1})+
c_\ell (\sigma^-_i \sigma^+_{i+1}- \hat{n}_i \hat{v}_{i+1} )\right] \\
\label{pr6b}
& & -  \left[
c_\ell (\sigma^+_m \sigma^-_{m+1} -  \hat{n}_m \hat{v}_{m+1})+
c_r (\sigma^-_m \sigma^+_{m+1} - \hat{v}_{m} \hat{n}_{m+1} )\right]
\eea
which we write in the form $\tilde{H}^{(m)} = \tilde{H}^{(m)}_b + B^{(m)}$ with bulk 
term $ \tilde{H}^{(m)}_b = \sum^\prime_i \tilde{h}^{b}_i$ given by
\eref{pr6a} transformation term $B^{(m)}$ given by \eref{pr6b}.

Using the properties of the particle creation and annihilation operators listed above
it is easy to check by straightforward computation
that  $h^{b}_i \ket{\mu^{N}_{z_1}} = 0$. On the other hand,
$ B^{(m)} \ket{\mu^{N}_{z_1}} = - (c_r-c_\ell) (\hat{n}_{m+1} - \hat{n}_m)  \ket{\mu^{N}_{z_1}}$.
This implies
$V_m B^{(m)} \ket{\mu^{N}_{z_1}} = - (c_r-c_\ell) (\hat{n}_{m+1} - \hat{n}_m) V_m \ket{\mu^{N}_{z_1}}$
since $V_m$ and the number operators $\hat{n}_i$ are both diagonal and hence commute.
With the projector property
$\hat{n}_k^2 = \hat{n}_k$ one has $q^{-2N/L} V_{m+1} = (1 + (q^{-2}-1) \hat{n}_{m+1} ) V_m$ and 
$q^{2N/L} V_{m-1} = (1 + (q^{2}-1) \hat{n}_m) V_m$. Therefore $c_r q^{-2N/L} V_{m+1}
+ c_\ell q^{2N/L} V_{m-1} - (c_r+c_\ell) V_m = - (c_r-c_\ell) (\hat{n}_{m+1} - \hat{n}_m) V_m$.
Putting these results together yields
\bel{Theorem1a}
\frac{\rmd}{\rmd t} \tilde{\mu}^{II,N}_{m}(t) = 
c_r q^{-\frac{2N}{L}} \tilde{\mu}^{II,N}_{m+1}(t) + c_l q^{\frac{2N}{L}} 
\tilde{\mu}^{II,N}_{m-1}(t) - (c_r+c_\ell) \tilde{\mu}^{II,N}_{m}(t)
\ee
for any $m\in \T_L$ and any $t\geq 0$.

Next we compute the time derivative of the normalization 
$R_m(t) = \bra{s} \rme^{-\tilde{H} t} \ket{ \mu^{II,N}_{m}}$. Using
\eref{Theorem1a} we get
\bea
\frac{\rmd}{\rmd t} R_m(t) & = & - \bra{s} \rme^{\tilde{H} t} \tilde{H} \ket{ \mu^{II,N}_{m}} \nonumber \\
& = & c_r q^{-2N/L} R_{m+1}(t) + c_\ell q^{2N/L} R_{m-1}(t) - (c_r+c_\ell) R_m(t) \\
& = & \left[ c_r q^{-2N/L}  + c_\ell q^{2N/L}  - (c_r+c_\ell) \right] R_m(t) \nonumber
\eea
where the last line follows from translation invariance. Therefore
\be
\mu^{II,N}_{k}(t) = \tilde{\mu}^{II,N}_{k}(t)/R(t) = 
\exp{[(-c_r q^{-2N/L}  - c_\ell q^{2N/L}  + c_r+c_\ell)t]}\tilde{\mu}^{II,N}_{k}(t)/R(0)
\ee
which implies the system of linear ODE's
\bel{ODE2}
\frac{\rmd}{\rmd t} \mu^{II,N}_{k}(t) = c_r q^{-2N/L} \mu^{II,N}_{k+1}(t) 
+ c_\ell q^{2N/L} \mu^{II,N}_{k-1}(t) -
\left( c_r q^{-2N/L}+ c_\ell q^{2N/L}\right) \mu^{II,N}_{k}(t).
\ee
Now it only remains to show that \eref{Theorem1b} satisfies this system of ODE's with the
initial condition $\mu^{II}_m(0) = \mu^{II}_m$.
It is elementary that the random walk transition probability of the theorem satisfies the
forward evolution equation 
\be
\frac{\rmd}{\rmd t}  p_t(l \iz m) = c_r q^{-2N/L} p_t(l-1 \iz m)
+ c_\ell q^{2N/L} p_t(l +1 \iz m) -
\left( c_r q^{-2N/L}+ c_\ell q^{2N/L}\right) p_t(l \iz m)
\ee
with initial condition $p_0(l \iz m) = \delta_{l,m}$.
The theorem thus follows from translation invariance $p_t(l +r \iz m + r) = p_t(l \iz m)$ 
$\forall r \in Z$ and periodicity $p_t(l + pL \iz m) = p_t(l \iz m)$ 
$\forall p \in Z$
of the transition probability. 
\qed

\begin{theo} 
\label{main2}
(Shock/Antishock evolution under local conditioning)
Let the initial distribution of the ASEP with $N$ particles be given
by the type I shock measure $\mu^{I,N}_{m}$ satisfying condition S
with shock at bond $m$ and antishock at bond $M$
and let $\tilde{\mu}^{I,N}_m(t)$ denote the unnormalized grandcanonically conditioned
measure at time $t$ under local conditioning at bond $M$ with weight $\rme^{s} = c_\ell/c_r$. 
Then we have  $\forall \, t\geq 0$
\bel{localcond}
\tilde{\mu}^{I,N}_m(t) = \rme^{(\tilde{c}_r + \tilde{c}_\ell - c_r - c_\ell)t}
\sum_{l=-M+1}^{M}p_t(l \iz m) \delta^{l-m} q^{\frac{2N(l-m)}{L}} \mu^{I,N}_{l}
\ee
with $p_t(l \iz m)$ defined in \eref{main} and $\delta=\frac{1+z_1}{1+q^2z_1}$.

\end{theo}

{\it Proof:} First we consider the evolution of unnormalized measures 
$\tilde{\mu}^{I,N}_m(t)$. By construction (described informally 
in Sec. 2) the matrix $\tilde{H}^M$ defined by \eref{pr5a} and boundary term $B_M$ 
given by \eref{pr5b} is the generator for the grandcanonically conditioned evolution 
under local 
conditioning at bond $(M,-M+1)$ with weight $\rme^{s} = c_\ell/c_r$. Therefore
$\rmd/\rmd t \ket{\tilde{\mu}^{I,N}_m} = - \tilde{H}^M \ket{\tilde{\mu}^{I,N}_m}$.
Notice next that $U= W V_M$ where $U$ is defined
in \eref{U} and $V_M$ is defined in \eref{pr2}. Therefore we may write $\tilde{H}^M = U^{-1} \tilde{H} U$ and we
have by
Theorem \eref{main} in conjunction with Lemma \eref{Lemma2}

\bea
- \tilde{H}^M \ket{\tilde{\mu}^{I,N}_m} 
& = & - Y_m q^{2 \frac{M-m}{L} N} U^{-1} \tilde{H} \ket{\tilde{\mu}^{II,N}_m}(t) \nonumber \\
& = & Y_m q^{2 \frac{M-m}{L} N} \left[ c_r q^{- \frac{2N}{L}} \frac{1}{Y_{m+1}} q^{- 2 \frac{M-m-1}{L} N}
\ket{\tilde{\mu}^{I,N}_{m+1}} \right. \nonumber \\
& & + c_\ell q^{ \frac{2N}{L}} \frac{1}{Y_{m-1}} q^{- 2 \frac{M-m+1}{L} N}
\ket{\tilde{\mu}^{I,N}_{m-1}}  \nonumber \\
& & \left. -(c_r + c_\ell)  \frac{1}{Y_{m}} q^{- 2 \frac{M-m}{L} N} \ket{\tilde{\mu}^{I,N}_m} \right] \quad (-M \leq m < M).
\eea

From \eref{Ym} we have $Y_m/Y_{m+1} = (1+z_1)/(1+z_2)$ and 
therefore
\bel{main2a}
\frac{\rmd}{\rmd t} \tilde{\mu}^{I,N}_{m}(t) = 
c_r \delta \tilde{\mu}^{I,N}_{m+1}(t) + c_\ell \delta^{-1} \tilde{\mu}^{I,N}_{m-1}(t) - 
(c_r+c_l) \tilde{\mu}^{I,N}_{m}(t)
\quad  \forall m \in \T \setminus {\{-M+1,M\}}.
\ee

For $m=M$ we have
\bea
- \tilde{H}^M \ket{\tilde{\mu}^{I,N}_M}(t) & = & - Y_M U^{-1} \tilde{H} \ket{\tilde{\mu}^{II,N}_M}(t) \nonumber \\
& = & Y_M  \left[ c_r q^{- \frac{2N}{L}} \frac{1}{Y_{-M+1}} q^{- 2 \frac{-1}{L} N}
\ket{\tilde{\mu}^{I,N}_{-M+1}} \right. \nonumber \\
& & + c_\ell q^{ \frac{2N}{L}} \frac{1}{Y_{M-1}} q^{- 2 \frac{1}{L} N}
\ket{\tilde{\mu}^{I,N}_{m-1}}  \nonumber \\
& & \left. -(c_r + c_\ell)  \frac{1}{Y_{m}}  \ket{\tilde{\mu}^{I,N}_m} \right].
\eea
Straightforward computation yields $Y_M/Y_{-M+1} = [(1+z_2)/(1+z_1)]^{L-1}$. 

For $m=-M$ we use Lemma \eref{Bernoullitrafo}.
This yields 
\bea
\label{main2b}
 \frac{\rmd}{\rmd t} \tilde{\mu}^{I,N}_{M}(t) & = & c_r \delta \delta^{-L} q^{-2N}
\tilde{\mu}^{I,N}_{-M+1}(t) + c_\ell \delta^{-1} \tilde{\mu}^{I,N}_{M-1}(t) - 
(c_r+c_l) \tilde{\mu}^{I,N}_{M}(t)  \\
\label{main2c}
 \frac{\rmd}{\rmd t} \tilde{\mu}^{I,N}_{-M+1}(t) & = & c_r \delta 
\tilde{\mu}^{I,N}_{-M+2}(t) + c_\ell \delta^{-1} \delta^{L} q^{2N} \tilde{\mu}^{I,N}_{M}(t) - 
(c_r+c_l) \tilde{\mu}^{I,N}_{-M+1}(t) .
\eea

From \eref{main2a} - \eref{main2c} one finds that he transformed measure 
$\ket{\tilde{\nu}^{I,N}_m}(t) = \delta^m q^{2Nm/L} \ket{\tilde{\mu}^{I,N}_m}(t)$ satisfies the system of 
linear ODE's \eref{Theorem1a}. Hence, by the same arguments as used in
\eref{main} we obtain \eref{localcond}.
\qed

\begin{cor}
\label{cornormalized}
The normalized grandcanonically conditioned
measure at time $t$ of Theorem \eref{main2} $\mu^{I,N}_m(t)$ is of the form
\bel{localcondnorm}
\mu^{I,N}_m(t) = \frac{1}{C(t)}
\sum_{l=-M+1}^{M}p_t(l \iz m) \delta^{l-m} q^{\frac{2N(l-m)}{L}} \mu^{I,N}_{l}
\ee
with a normalization constant $C(t) = \sum_{l=-M+1}^{M}p_t(l \iz m) \delta^{l-m} q^{\frac{2N(l-m)}{L}}
\inprod{s}{\mu^{I,N}_{l}}$ that
is strictly positive and finite for all $t\geq 0$.
In particular,
\bel{localcondnormstat}
\mu^{I^\ast,N}_m := \lim_{t\to \infty} \mu^{I,N}_m(t) =
\frac{1}{C^\ast}
\sum_{l=-M+1}^{M} \delta^{l-m} q^{\frac{2N(l-m)}{L}} \mu^{I,N}_{l}
\ee
with $C^\ast =  \sum_{l=-M+1}^{M} \delta^{l-m} q^{\frac{2N(l-m)}{L}}
\inprod{s}{\mu^{I,N}_{l}}$ is the unique stationary conditioned limiting measure.
\end{cor}

The properties of $C(t)$ are obvious from the fact it is a transition probability for a random walk with unnormalized
positive initial distribution and that the random walk propagator on $\T_L$ satisfies $\lim_{t\to \infty} p_t(l \iz m)
=1/L$ $\forall l,m \in \T_L$. The uniqueness of the measure follows from the Perron-Frobenius theorem for the
weighted generator. 
\qed

\section{Final remarks}

Our results are exact results on the lattice scale and for finite times. With regard to the
macroscopic large-deviation theory of \cite{Bodi05} 
Theorem \eref{main} suggests that there exists a value of $s(\nu)$ such that
the optimal macroscopic density profile obtained from the 
large-deviation theory of \cite{Bodi05} in the limit $\nu \to \infty$ is of the form 
$\rho(x,t) = 1/2 [ 1- \tanh{\left(E(x+a-vt)\right)} ]$ with speed $v = c_r \exp{(-2\rho s)} - c_\ell \exp{(2\rho s)}$
and a jump discontinuity at some point $x^\ast(t)$ that also travel with speed $v$. 
The stationary solution in Theorem  \eref{main2}  for local conditioning with weight $\rme^{s} = c_\ell/c_r$ 
indicates the existence of an antishock that remains microscopically sharp and does not fluctuate.
This behaviour is in sharp contrast to the typical behavior of the ASEP where an 
antishock evolves into a rarefaction wave.

The evolution of the shock initial measures as a one-particle random walk indicates a self-duality
relation of the conditioned ASEP with periodic boundary conditions, which is absent for the unconditioned
dynamics.
Results from \cite{Beli02} for the microscopic dynamics of more than one shock, each satisfying
condition S, and properties of the underlying representation theory of the quantum group
$U_q[SU(2)]$ for periodic systems \cite{Pasq90}
suggests that one may study the behaviour of shock/antishock measures with several
shocks under global and local conditioning. One then expects the evolution of $n$ shocks 
to be given by the dynamics of $n$ interacting
particles. The results of \cite{Imam11} might then 
provide information about more general solutions of the large deviation theory and about their
fluctuations.

\section*{Acknowledgements}
GMS thanks FAPESP for financial support and IME at the University of S\~ao Paulo for kind hospitality. 
VB acknowledges financial support of FAPESP and CNPq.

\end{document}